\documentclass[a4paper]{article}

\usepackage[british]{babel}
\usepackage[utf8]{inputenc}
\usepackage[T1]{fontenc}
\usepackage{graphicx}
\usepackage{amssymb}
\usepackage{amsthm}
\usepackage{mathrsfs}
\usepackage{amsmath}
\usepackage{subfigure}
\usepackage{authblk}
\usepackage[sort&compress,square,comma,authoryear]{natbib}

\usepackage[
  colorlinks=false,
  pdfborder={0 0 0},
  pdfauthor={Emmanuel Roubin, Jean-Baptiste Colliat},
  pdftitle={Analytical link between structural strength size effect and material random heterogeneity},
  pdfsubject={Material science, Statistical mechanics},
  pdfkeywords={Size effect, Theoretical scaling law, Excursion sets, correlated Random Fields, Euler characteristic},
  pdfproducer={LaTeX with hyperref package},
  pdfcreator={pdflatex},
  breaklinks=true
]{hyperref}

\usepackage{tikz,pgfplots}
\pgfplotsset{every tick label/.append style={font=\scriptsize},
             legend                   style={font=\scriptsize},
             label                    style={font=\scriptsize}}

\pgfplotsset{
  legend entry/.initial=, every axis plot post/.code={%
    \pgfkeysgetvalue{/pgfplots/legend entry}\tempValue
    \ifx\tempValue\empty
    \pgfkeysalso{/pgfplots/forget plot}%
    \else
    \expandafter\addlegendentry\expandafter{\tempValue}%
    \fi
  },
}

\pgfplotsset{
        width =    \textwidth/1.5,
        height=0.75\textwidth/2
}

\usetikzlibrary{shapes.misc}
\tikzset{cross/.style={cross out, draw=black, minimum size=2*(#1-\pgflinewidth), inner sep=0pt, outer sep=0pt, thick=2}, cross/.default={2pt}}

\definecolor{myblue}{rgb}{0.00,0.39,0.67}
\definecolor{myred}{rgb}{0.86,0.08,0.14}
\definecolor{mygreen}{rgb}{0.07,0.52,0.07}
\definecolor{myorange}{rgb}{0.99,0.32,0.08}
\definecolor{mygray}{rgb}{0.5,0.5,0.5}

\definecolor{colorA}{rgb}{0.00,0.39,0.67}
\definecolor{colorB}{rgb}{0.86,0.08,0.14}
\definecolor{colorC}{rgb}{0.07,0.52,0.07}
\definecolor{colorD}{rgb}{0.99,0.32,0.08}

\newcommand\markA{triangle*}
\newcommand\markB{otimes*}
\newcommand\markC{diamond*}

\def\ie {\textit{i.e.}~}

\def\etc{\textit{etc}}

\def\x {\boldsymbol{x} }
\def\y {\boldsymbol{y} }

\def\s { \sigma }

\def\slim { \s_{\text{y}} }
\def\sfail { \s_{\text{f}} }
\def\sfailbis { \tilde{\s}_{\text{f}} }
\def\ps{p_\text{safe}}
\def\pf{p_\text{fail}}

\def\exp{ \text{exp} }

\def\R3{ \mathbb{R}^3 }

\def\max{ \text{max} }

\def\HS{ H_s }

\def\ES{ E_s }

\def\lset { \s }

\def\lsetmax { \s_\max }
\def\lsetlow { K_\text{low} }
\def\lfactor { \gamma }
\def\sublset { g(\lset) }

\def\zlset{[0,\lset]}

\def\w{ \omega }

\def\GRF{ g }

\def\GRRF{ \GRF_r }

\def\MRF{ M }

\def\normal{ \mathscr{N} }
\def\lognormal{ \text{Log}\!-\!\normal }
\def\Proba{ P }
\def\Expec{ \mathbb{E} }

\def\std{ s }
\def\var{ \std^2 }
\def\cvar{ c }
\def\logstd{ \std_\text{log} }
\def\logvar{ \std_\text{log}^2 }
\def\logcvar{ \cvar_\text{log} }
\def\esp{ \mu }
\def\logesp{ \mu_\text{log} }
\def\Lc{ l_c }
\def\Msize { a }

\def\rlinv {\beta}
\def\EC{ \chi }

\def\Size{ l }
\def\SpS{ \Size_f }

\def\cov{ \mathcal{C} }

\def\wbull{ k }

\begin{document}

\title{Analytical link between structural strength size effect and material random heterogeneity}

\author[a]{Emmanuel Roubin\thanks{Corresponding author: \href{mailto:emmanuel.roubin@3sr-grenoble.fr}{emmanuel.roubin@3sr-grenoble.fr}}}
\author[b]{Jean-Baptiste Colliat}

\date{}
\setcounter{Maxaffil}{0}
\renewcommand\Affilfont{\itshape\small}

\affil[a]{Univ. Grenoble Alpes, CNRS, Grenoble INP\thanks{Institute of Engineering Univ. Grenoble Alpes}, 3SR, F-38000 Grenoble, France}
\affil[b]{Laboratoire de m\'ecanique multiphysique multi\'echelle, Universit\'e de Lille, CNRS, Centrale Lille\\ 59000 Lille, France.}

% \cortext[cor]{}

\maketitle

\begin{abstract}
    A theoretical scaling law for the size effect of the strength of brittle materials is presented. To some extend, it can be seen as an extension of the well known Weibull law. For that a correlated Random Fields is used to model the heterogeneities of the material. Thanks to recent results on the geometry of excursion sets, one can analytically compute the whole probability distribution function for the strength of a structure of a given size. Then, using this PDF, the structural strength associated to any failure probability can be derived.
\end{abstract}

\section{Introduction\label{sec:int}}
\par Size effects related to the strength of heterogeneous materials are a subject of major interest for more than three decades. Still, there are multiple alternatives to provide scaling laws, most of them being based on mechanical considerations.
The seminal results come from the early studies of \citep{weibull_statistical_1951} based on the theory of the weakest link.
The authors proposed an analytical solution for the structural failure probability, considering a set of independent brittle links with a given probability of local failure.
With no spatial correlation between each link, this theory leads implicitly to size effects at large scale.
More recently, the two current theories of Z.P.~Ba\v{z}ant and A.~Carpenteri, trying to describe the size effect for a broader range of scales and materials, are the main results of the extensive literature existing on this topic.
The former tends, in many ways, to describe the size effect using both non-local model and stochastic approach \citep{sab_unified_1993}, or more recently using the so-called energetic-statistical size effect mixing strength redistribution theory in a fracture process zone and Weibull's theory \citep{bazant_probability_2004}.
The latter considers material heterogeneities with a fractal model in order to represent size effects for quasi-brittle materials \citep{carpinteri_mechanics_2003}.
Finally, numerical simulations have been made using stochastic integrations and correlated Random Fields in order to describe material properties \citep{colliat_stochastic_2007,gregoire_failure_2013}.
These methods are time consuming and the underlying numerical implementation brings an inevitable limitation regarding the observation scale.

\par Following \citep{carpinteri_are_2005} we think that the heterogeneous geometry at fine scale is  of major importance to explain those effects.
Hence we propose to use \emph{correlated Random Fields} to assess the representation of the heterogeneous aspect of materials.
In addition to the usual characteristics of Random Variables (mean, variance, \etc), correlated Random Fields have a spatial structure that can be statistically controlled through their underlying \emph{covariance functions}.
Typically, assuming the isotropy of the material, the latter may be defined in terms of the so-called \emph{correlation length}.
Several aspects of this spatial structure, such as the expected number of upcrossings or the expected distance between maxima, correspond to morphological parameters which allows us quantify the scale of observation (structural scale) in comparison with the scale of the heterogeneities (material scale).
On the one hand, dealing with small structures, the correlation length may be comparable (or even larger) to the specimen size, thus leading to Random Fields realizations that are almost constant in space (but still random).
On the other hand, when considering large structures, the ratio between the correlation length and the specimen size is driven to zero.
Hence each realisation of such fields can be seen, in the limit, as a white noise.
In-between, we show that the use of correlated Random Fields leads to a continuous and highly nonlinear evolution of the strength along the specimen (or structure) size.

\par In this study, a theoretical method to describe strength size effects for brittle heterogeneous materials is proposed.
It extends the Weibull theory to a wider range of scales by exploiting the spatial structure of a correlated Random Fields in the representation a local failure stress.
A continuum representation of this spatial variability through scales is theoretically made by controling the ratio between the size of the Random Field domaine of definition and its correlation length.
The cornerstone of this method is to benefit from a theoretical result from \citep{adler_geometry_1981} that links the expected topology (\ie the Euler Characteristic) of the Random Field excursion to the exceedance probability and thus to the failure probability of the structure.
This theoretical relationship leads to a purely analytical model where, contrary to stochastic integration methods, the simulation of a high number of realisations is not necessary.
Hence, there is no scale limitation.

% \par The outline is as follows: Section~\ref{sec:A} introduces the underlying principle of excursion sets. For that matter, a brief review on correlated Random Fields is first made, then results on the expected Euler characteristic of excursions are given. Section~\ref{sec:B} provides the analytical model for strength size effect of brittle materials, using the fact that the probability that the global maximum of a Random Field exceeds a given threshold is well approximated by the expected Euler characteristic of the excursion set defined for the same threshold (provided the threshold is ``high'' or ``low'') \citep{adler_random_2007}.
%\par It is worth noting here that some recent results \citep{adler_random_2007} give the expectations of the geometrical (e.g volume and surface) and topological (Euler characteristics) measures of these excursions. Those results, coming from  thus providing a predictable morphological tool (statistically speaking). It is this latter approach that is explored in this paper.
%In this section another approach using expected values of excursion set characteristics is proposed. Up to this point, this mathematical tool has been used in order to yield actual morphologies and predict their geometrical and topological characteristics. The continuum aspect of correlated Random Fields was concealed by the level set method used which led to discrete fields (excursion sets). Herein, Random Fields directly represent a material property (ultimate stress field) and the excursion set theory is used in order to catch statistical information on its extrema.

\section{Material random properties modelling\label{sec:A}}

\par Correlated Random Fields are very efficients tools in order to represent the random aspect of heterogeneous materials.
They can be used according to two very different ways.
Firstly, their values can directly define any mechanical or physical property, thus leading to a continuum representation of the heterogeneous aspect of a media.
Combined with a stochastic integration method (such as classical Monte-Carlo integration \citep{larrard_influence_2012} or Spectral Stochastic Finite Element \citep{matthies_galerkin_2005}) Random Fields are thus a very convenient way to model parametric uncertainties.
Secondly, by explicitly defining the physical boundaries of the heterogeneities as a level set of a realisation of a Random Field, the domain of which can be divided into several subdomains, referred to as excursion sets \citep{roubin_meso-scale_2015}.

\par In this study, correlated Random Fields are directly used to define the tensile strenght but, the excursion sets theory is used to predict statistical properties of the mechanical response.

\subsection{The correlation length as a scale ratio}
\par We call scale ratio (noted $\rlinv$) the ratio between the specimen and the heterogeneities caracteristical size.
% This ratio can directly be expressed from the correlated Random Field definition.
Even though more complex distributions can be used, herein, for sake of simplicity, correlated Random Fields $\GRF(\x,\w)$ are defined over a parameter space $\MRF$ as isotropic, stationary fields with Gaussian $\normal(\esp,\var)$ or Gaussian related distribution and Gaussian covariance function $\cov$ defined by:
\begin{equation}
    \cov(\|\x-\y\|) = \var e^{-\|\x-\y\|^2/\Lc^2},
  \label{eq:covariance}
\end{equation}
where $\Lc$ is the so called correlation length.
% As it has been pointed out during the introduction, the correlation length $\Lc$ can been seen as a scale parameter.
% Herein, a constant value of $\Lc$ is chosen, setting the average size of the heterogeneities considered.

\par The size of the domain $\MRF$ where the Random Field is defined represents the size of the whole structure.
If $\Msize$ is the characteristic length of $\MRF$ (for example: the length of a segment in the one-dimensional space, the length of the side of a square in a two-dimensional space\dots).
In order to let the heterogeneity size unspecified, the dimensionless ratio
\begin{equation}
  \rlinv = \frac{a}{\Lc}
  \label{eq:scale-ratio}
\end{equation}
 is taken into consideration, its value determining the observation scale.

\begin{description}
  \item [For $\rlinv \ll 1$] the structure is very small compared to the heterogeneity size. The Random Field tends to be a constant field and is equivalent to a simple Random Variable with no spatial variation.
  It clearly represents the material scale and the validity range of Continuum Damage Mechanics (CDM), for which the failure stress does not depend on the size of the structure.
  \item [For $\rlinv \gg 1$] the structure is very large compared to the heterogeneity size.
  The Random Field tends to be a white noise (completely uncorrelated), leading to a loss of spatial structures.
  It represents the case of large structures corresponding to the Weibull theory.
  \item [For $\rlinv \approx 1$] The Random Field represents the missing scale range where the continuum statistical information of correlated Random Fields for various $\rlinv$ can link together material and large structure scales.
\end{description}

\subsection{Probabilistic definition of a one dimensional failure criterion}
\par A one dimensional structure in tension is considered.
The material failure criterion, which is the source of uncertainty, is defined by a correlated Random Field $\slim(x,\w)$ with correlation length $\Lc$.
Due to the positiveness of material tensile strength, the log-normal distribution $\lognormal(\esp,\var)$ (which is Gaussian related by the exponential function) is used.
The field is defined over a one-dimensional bar $\MRF$ of size $\Msize$.

\par The structural failure of $\MRF$ occurs when the stress field (which is constant in tension) reaches the minimum value of $\slim(x,\w)$.
The most intuitive way of defining the structural failure stress, noted $\sfailbis(\w)$, is through a definition for a given realisation $i$ of $\slim$:

\begin{equation}
  \sfailbis(\w_i) = \underset{x \in \MRF}{\inf}\left(\slim(x,\w_i)\right)
  \label{eq:size-effect:failure-criterion}
\end{equation}

\par
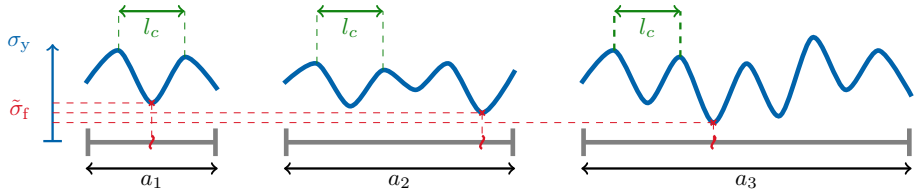
\begin{figure*}[h!]
  \centering
  \resizebox{\textwidth}{!}{  \begin{tikzpicture}

    \draw[|->,myblue,line width=1] (-0.5,0) -- (-0.5,1.5);
    \draw[myblue] (-1,1.5) node {$\slim$};
    \draw[myred] (-1,0.5) node {$\sfailbis$};
    
    \newcommand\xshiftA{7.5}
    \begin{scope}[shift={(\xshiftA,0)}]    
    % BEAM A --- beam
    \draw[mygray,|-|,line width=2] (0,0) -- (5,0);
    \draw[<->,line width=1] (0,-0.4) -- (5,-0.4);
    \draw (2.5,-0.6) node {$\Msize_3$};                   
    % BEAM A --- sigma_y
    \draw[myblue,line width=2] plot [smooth] coordinates {
      (0.0,0.9)
      (0.5,1.4)
      (1.0,0.6)
      (1.5,1.3)
      (2.0,0.3)
      (2.5,1.2)
      (3.0,0.4)
      (3.5,1.6)
      (4.0,0.8)
      (4.5,1.4)
      (5.0,0.7)};
    \newcommand\XfA{2.0}
    \newcommand\YfA{0.3}
    % BEAM A --- correlation length
    \draw[mygreen,dashed,line width=1] (0.5,1.4) -- (0.5,2);
    \draw[mygreen,dashed,line width=1] (1.5,1.3) -- (1.5,2);
    \draw[mygreen,<->,line width=1] (1.5,2) -- (0.5,2);
    \draw[mygreen] (1,1.75) node {$\Lc$};
    % BEAM A --- sigma_f
    \draw (\XfA,\YfA) node[cross,myred] {};
    \draw[myred,line width=1] (\XfA,-0.1) .. controls (\XfA-0.1,-0.05) and (\XfA+0.1,0.05) .. (\XfA,0.1);
    \draw[-,dashed,myred] (-0.5-\xshiftA,\YfA) -- (\XfA,\YfA);
    \draw[-,dashed,myred] (\XfA,\YfA) -- (\XfA,0);
    \end{scope}

    \newcommand\xshiftB{3.0}
    \begin{scope}[shift={(\xshiftB,0)}]
    % BEAM B --- beam
    \draw[mygray,|-|,line width=2,] (0,0) -- (3.5,0);
    \draw[<->,line width=1] (0,-0.4) -- (3.5,-0.4);
    \draw (1.75,-0.6) node {$\Msize_2$};
    % BEAM B --- sigma_y
    \draw[myblue,line width=2] plot [smooth] coordinates {
      (0.0,0.9)
      (0.5,1.2)
      (1.0,0.55)
      (1.5,1.1)
      (2.0,0.8)
      (2.5,1.2)
      (3.0,0.45)
      (3.5,1.1)};
    \newcommand\XfB{3.0}
    \newcommand\YfB{0.45}
    % BEAM B --- correlation length
    \draw[mygreen,dashed] (0.5,1.2) -- (0.5,2);
    \draw[mygreen,dashed] (1.5,1.1) -- (1.5,2);
    \draw[mygreen,<->,line width=1] (1.5,2) -- (0.5,2);
    \draw[mygreen] (1,1.75) node {$\Lc$};
    % BEAM B --- sigma_f
    \draw (\XfB,\YfB) node[cross,myred] {};
    \draw[myred,line width=1] (\XfB,-0.1) .. controls (\XfB-0.1,-0.05) and (\XfB+0.1,0.05) .. (\XfB,0.1);
    \draw[-,dashed,myred] (-0.5-\xshiftB,\YfB) -- (\XfB,\YfB);
    \draw[-,dashed,myred] (\XfB,\YfB) -- (\XfB,0);
    \end{scope}.

    \newcommand\xshiftC{0}
    \begin{scope}[shift={(\xshiftC,0)}]
    % BEAM C --- beam
    \draw[mygray,|-|,line width=2] (0,0) -- (2,0);
    \draw[<->,line width=1] (0,-0.4) -- (2,-0.4);
    \draw (1,-0.6) node {$\Msize_1$};
    % BEAM C --- sigma_y
    \draw[myblue,line width=2] plot [smooth] coordinates {
      (0.0,0.9)
      (0.5,1.4)
      (1.0,0.6)
      (1.5,1.3)
      (2.0,0.8)};
    \newcommand\XfC{1.0}
    \newcommand\YfC{0.6}        
    % BEAM C --- correlation length
    \draw[mygreen,dashed] (0.5,1.4) -- (0.5,2);
    \draw[mygreen,dashed] (1.5,1.3) -- (1.5,2);
    \draw[mygreen,<->,line width=1] (1.5,2) -- (0.5,2);
    \draw[mygreen] (1,1.75) node {$\Lc$};
    % BEAM C --- sigma_f
    \draw (\XfC,\YfC) node[cross,myred] {};
    \draw[myred,line width=1] (\XfC,-0.1) .. controls (\XfC-0.1,-0.05) and (\XfC+0.1,0.05) .. (\XfC,0.1);
    \draw[-,dashed,myred] (-0.5-\xshiftC,\YfC) -- (\XfC,\YfC);
    \draw[-,dashed,myred] (\XfC,\YfC) -- (\XfC,0);
    \end{scope}
  \end{tikzpicture}}
  \caption{Illustration of the material failure stress repartition on bars of various sizes.\label{fig:size-effect:1D-various-sizes}}
\end{figure*}

 The correlation length $\Lc$, because being fixed by the heterogeneity size, is set to be the same for all test.
Thus, the different scales are represented by defining the material failure stress $\slim(x,\w)$ by a unique covariance function but over various sizes $\Msize$ of the bar $\MRF$, as represented in Figure~\ref{fig:size-effect:1D-various-sizes}.
This Figure also represents the structural failure criterion as the minimum of the material failure criterion (as defined in Equation~\eqref{eq:size-effect:failure-criterion}).

\par The limitation of Equation~\eqref{eq:size-effect:failure-criterion} is that the failure criterion $\sfailbis$ is defined as a random variable.
A reformulation as a full distribution in terms of safety probability\footnote{This probability can be seen as the complementary of the failure probability $\pf$ (probability that the structure fails), giving the relation: $\ps=1-\pf$.} $\ps$ leads to a more general definition:
\begin{equation}
  \label{eq:size-effect:global-stress}
  \sfail(\ps) = \left\{ \s \ | \ \Proba\left\{ \underset{x\in \MRF}{\inf}\left( \slim(x,\w) \right) \leq \s \right\} = 1-\ps \right\},
\end{equation}
where $\Proba\{ \inf( \slim(x,\w) ) \leq \s \}$ is the probability that the minimal value of $\slim$ over the bar is smaller than a given stress state $\s$.
\par In the next section, we show how the excursion sets theory is used in order to obtain an analytical knowledge of this probability which, in turns, gives a analytical knowledge of $\sfail(\ps)$.

\subsection{Failure interpretated as an excursion and its Euler caracteristic}
\par In order to have an analytical definition of the probability mentioned just above, $\Proba\{ \inf( \slim(x,\w) ) \leq \s \}$, results from the excursion set theory of correlated Random Field are used.
The key point is that Equation~\eqref{eq:size-effect:global-stress} can directly be linked with the Euler characteristic\footnote{The Euler characteristic of a one dimensional set is simply its number of connected components.} $\EC$ of the excursion set $\ES$ defined by:
\begin{equation}
  \ES(\s) = \left\{x \in \MRF \ | \ \slim(x,\omega)\leq\s \ \right\}.
\end{equation}
\par
\begin{figure}[h!]
  \centering
  \subfigure[``Low'' threshold $\s$ \label{graph:excursion:1D:low}]  {\begin{tikzpicture}[scale=1.0]
  \newcommand\xscale{1.5}

  % axes
  \draw[->,line width=1,dotted] (-0.2*\xscale,0) -- (5*\xscale,0);
  \draw (5.3*\xscale,0) node {$x$};
  %\draw[->,line width=1,dotted] (0*\xscale,-0.2) -- (0*\xscale,2.5);
  \draw[|->,line width=1,myblue] (0*\xscale,-0.025) -- (0*\xscale,2.5);
  \draw[myblue] (-0.4*\xscale,2.4) node {$\slim$};
  
  % Beam M
  \draw[mygray,|-|,line width=2] (0.5*\xscale,0) -- (4.5*\xscale,0);
  \draw[mygray] (4*\xscale,-0.4) node {$\MRF$};

  % Random Field
  %\draw[myblue] (4*\xscale,1.5) node {$\slim$};
  \draw[myblue,line width=2] plot [smooth,tension=0.7] coordinates {(0.5*\xscale,1.5)(0.75*\xscale,1.7)(1.3*\xscale,1.1)(1.9*\xscale,2.0)(2.7*\xscale,0.7)(3.4*\xscale,2.3)(3.8*\xscale,1.8)(4.3*\xscale,2.0)(4.5*\xscale,1.8)};

  % Threshold
  \newcommand\ylset{0.9}
  \draw[myred] (-0.5*\xscale,\ylset) node {$\s$};
  \draw[myred,line width=1,dashed] (-0.2*\xscale,\ylset) -- (5*\xscale,\ylset);

  % hitting set
  %\draw[-|,line width=2] (0*\xscale,0) -- (0*\xscale,\ylset);
  %\draw (-0.5*\xscale,0.4) node {$\HS$};
  
  % Excrusion
  \newcommand\xHSAA{2.45}
  \newcommand\xHSAB{2.9}
  \draw[mygreen] (2.7*\xscale,-0.5) node {$\ES(\s)$};
  \draw[mygreen,|-|,line width=2] (\xHSAA*\xscale,0) -- (\xHSAB*\xscale,0);
  \draw[mygreen,dashed,line width=1] (\xHSAA*\xscale,0) -- (\xHSAA*\xscale,\ylset);
  \draw[mygreen,dashed,line width=1] (\xHSAB*\xscale,0) -- (\xHSAB*\xscale,\ylset);  
\end{tikzpicture}}
  \subfigure[``High'' threshold $\s$ \label{graph:excursion:1D:high}]{\begin{tikzpicture}[scale=1.0]
  \newcommand\xscale{1.5}

  % axes
  \draw[->,line width=1,dotted] (-0.2*\xscale,0) -- (5*\xscale,0);
  \draw (5.3*\xscale,0) node {$x$};
  %\draw[->,line width=1,dotted] (0*\xscale,-0.2) -- (0*\xscale,2.5);
  \draw[|->,line width=1,myblue] (0*\xscale,-0.025) -- (0*\xscale,2.5);
  \draw[myblue] (-0.4*\xscale,2.4) node {$\slim$};

  % Beam M
  \draw[mygray,|-|,line width=2] (0.5*\xscale,0) -- (4.5*\xscale,0);
  \draw[mygray] (4*\xscale,-0.4) node {$\MRF$};

  % Random Field
  %\draw[myblue] (4*\xscale,1.5) node {$\slim$};
  \draw[myblue,line width=2] plot [smooth,tension=0.7] coordinates {(0.5*\xscale,1.5)(0.75*\xscale,1.7)(1.3*\xscale,1.1)(1.9*\xscale,2.0)(2.7*\xscale,0.7)(3.4*\xscale,2.3)(3.8*\xscale,1.8)(4.3*\xscale,2.0)(4.5*\xscale,1.8)};

  % Threshold
  \newcommand\ylset{1.3}
  \draw[myred] (-0.5*\xscale,\ylset) node {$\s$};
  \draw[myred,line width=1,dashed] (-0.2*\xscale,\ylset) -- (5*\xscale,\ylset);

  % hitting set
  %\draw[-|,line width=2] (0*\xscale,0) -- (0*\xscale,\ylset);
  %\draw (-0.5*\xscale,0.4) node {$\HS$};
  
  % Excrusion A
  \newcommand\xHSAA{2.3}
  \newcommand\xHSAB{3.05}
  \draw[mygreen,|-|,line width=2] (\xHSAA*\xscale,0) -- (\xHSAB*\xscale,0);
  \draw[mygreen,dashed,line width=1] (\xHSAA*\xscale,0) -- (\xHSAA*\xscale,\ylset);
  \draw[mygreen,dashed,line width=1] (\xHSAB*\xscale,0) -- (\xHSAB*\xscale,\ylset);  

  % Excrusion B
  \newcommand\xHSBA{1.05}
  \newcommand\xHSBB{1.45}
  \draw[mygreen,|-|,line width=2] (\xHSBA*\xscale,0) -- (\xHSBB*\xscale,0);
  \draw[mygreen,dashed,line width=1] (\xHSBA*\xscale,0) -- (\xHSBA*\xscale,\ylset);
  \draw[mygreen,dashed,line width=1] (\xHSBB*\xscale,0) -- (\xHSBB*\xscale,\ylset);

  \draw[mygreen] (1.95*\xscale,-0.5) node {$\ES(\s)$};
  \draw[mygreen,->,line width=1] (1.4*\xscale,-0.5) -- (1.2*\xscale,-0.2);
  \draw[mygreen,->,line width=1] (2.5*\xscale,-0.5) -- (2.7*\xscale,-0.2);
  
\end{tikzpicture}}
  \caption{Representation of a one-dimensional excursion sets.\label{graph:excursion:1D}}
\end{figure}
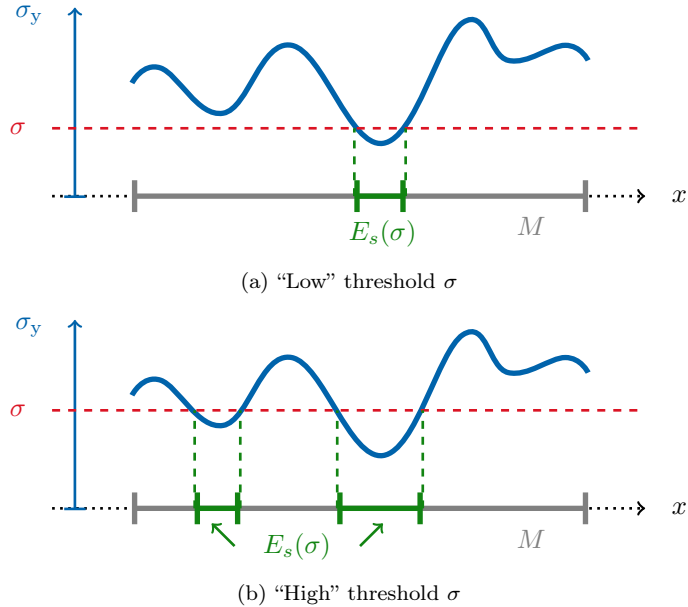

 As shown in Figure~\ref{graph:excursion:1D}, the excursion setis a sub-domain of $\MRF$ where the stress state $\s$ is greater than the material failure criterion $\slim(x,\w)$, $\s$ playing the role of threshold\footnote{It is for this reason that, within the excursion set framework, $\s$ is also refered to as threshold.}.
The previous structural failure stress $\sfailbis(\w_i)$ can now be seen in terms of excursion set, $\sfail$ being the stress state when, with increasing $\s$, $\ES(\s)$ changes from being a void subset of $\MRF$ ($\EC = 0$) to a single connected component ($\EC = 1$).

\par One of the results of \citep{adler_new_2008} makes the link between excursion set theory (through the expected value of the Euler characteristic) and the probability of reaching the minima of the underlying Random Fields.
In the present case, it gives a new formulation of the failing probability~\eqref{eq:size-effect:global-stress}:
\begin{equation}
  \label{eq:size-effect:link-ECandP}
  \Proba\left\{ \underset{x\in \MRF}{\inf}\left( \slim(x,\w) \right) \leq \s \right\} \approx \Expec\left\{ \EC\left(\ES(\s)\right) \right\},
\end{equation}
for low $\s$.
Furthermore, \citep{adler_new_2008} gives an analytical link between the excursion sets parameters (Random Field $\slim$ and threshold $\s$) and the expected value of the Euler characteristic:
\begin{equation}
  \label{eq:size-effect:expected-EC}
  \Expec\left\{ \EC\left(\ES(\s)\right) \right\} = \frac{\rlinv}{\sqrt{2}\pi} e^{-\sublset^2} + \frac{1}{\sqrt{\pi}}\int_{-\infty}^{\sublset} e^{-x^2} dx,
\end{equation}
where, for a log-normal distribution $\lognormal(\esp,\var)$:
\begin{equation}
  \sublset= \frac{\ln(\lset)-\esp}{\sqrt{2}\std}.
  \label{eq:substitution-log-normal}
\end{equation}
\par Replacing Equations~\eqref{eq:size-effect:link-ECandP} and \eqref{eq:size-effect:expected-EC} into Equation~\eqref{eq:size-effect:global-stress} gives a direct analytical knowledge of the failing probability, which is the main feature of the present model.

\par Before moving to the results and because it is usefull in order to understand the behavior of the probabilistic model, a simple analysis of Equation~\eqref{eq:size-effect:expected-EC} is proposed.

\subsection{Expected Euler characteristic and scale ratio}
\par
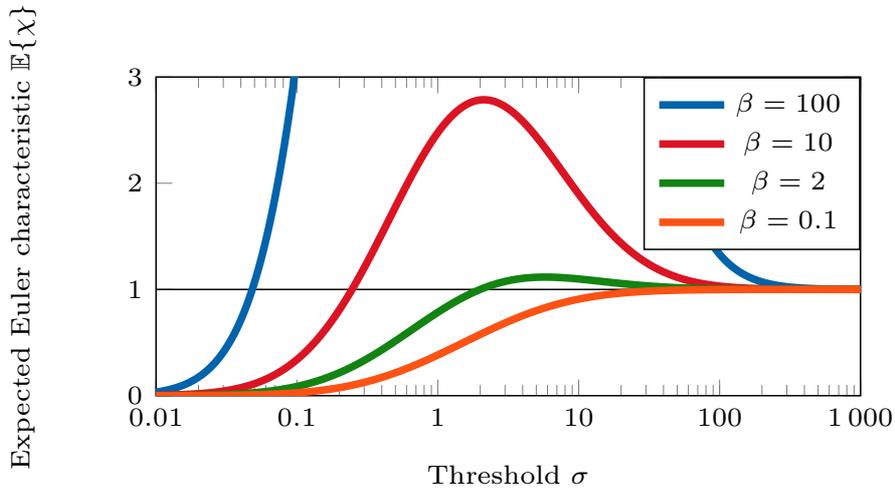
\begin{figure}[h!]
  \centering
  \resizebox{\textwidth}{!}{\begin{tikzpicture}
  \begin{semilogxaxis}[
      xlabel={Threshold $\lset$},
      ylabel={Expected Euler characteristic $\Expec\{\EC\}$},
      ymin=0.0,
      ymax=3.0,
      xmin=0.01,
      xmax=1000,
      log ticks with fixed point,
      x tick label style={/pgf/number format/1000 sep=\,},
      legend style={at={(1,1)},anchor=north east},
    ]

    \addplot[black,no markers] coordinates { (0.01,1) (1000,1) };
    \addplot[colorA, line width=2, legend entry={$\rlinv=100$}] table{tikz/data/lognormal_1D_Lc1_T100_th.dat};
    \addplot[colorB, line width=2, legend entry={$\rlinv=10$}]  table{tikz/data/lognormal_1D_Lc1_T10_th.dat};
    \addplot[colorC, line width=2, legend entry={$\rlinv=2$}]   table{tikz/data/lognormal_1D_Lc1_T2_th.dat};
    \addplot[colorD, line width=2, legend entry={$\rlinv=0.1$}] table{tikz/data/lognormal_1D_Lc1_T0.1_th.dat};
\end{semilogxaxis}

\end{tikzpicture}}
  \caption{Expected Euler characteristic $\Expec\{\EC\}$ of one dimensional excursion sets as a function of the threshold $\s$ for different length ratios $\rlinv$. The Gaussian related distribution is log-normal and is based on Gaussian correlated Random Field of mean $\esp=0.5$, variance $\var=2$ and correlation length $\Lc=1$ defined over a segment $\MRF$ of length $\Msize=100,10,2$ and $0.1$ ($\rlinv=100,10,2$ and $0.1$, respectively).\label{graph:ECs:1D}}
\end{figure}

 Considering the expected values of the Euler characteristic as a function of the stress state $\s$, Figure~\ref{graph:ECs:1D} shows the theoretical curves of Equation~\eqref{eq:size-effect:expected-EC} for various length ratios $\rlinv$.
\par The global behaviour of the Euler characteristic for different scale ratios can physically be understood.
The value of the maximum of each corresponds to the maximum number of disconnected components and, therefore, it is natural to see it decreases along with the scale ratio.
A scale ratio of $\rlinv=100$ gives a maximum of about $25$ components whereas $\rlinv=10$ gives a maximum around $2.7$ and $\rlinv=2$, a maximum of $1.1$.
For lower scale ratios the curve seems monotonic with no more maximum but actually, the maximum is very small for large values of the threshold $\s$.
When $\rlinv\rightarrow0$, the correlated Random Field tends to be constant over $\MRF$ (no fluctuations), $\s_\max\rightarrow\infty$ and the maximum $\EC$ is naturally $1$, the excursion being either empty or the full domain.
In this case, it can be understood that the expected value of the Euler characteristic corresponds to the probability of reaching the value of the Random Field, thus getting the full domain excursion.
Now, by decreasing $\rlinv$, the Random Field is not constant anymore but the same reasoning can be made for ``low'' thresholds, where the excursion set is either void or one single connected component.
Thus, for ``low'' thresholds, the Euler characteristic is directly linked with the probability of reaching the global minimum and reads, as in \citep{adler_new_2008} (Equation~\eqref{eq:size-effect:link-ECandP}).

\par In conclusion, even though $\sfail$ and $\sfailbis$ represents the same physical phenomena (and thus have the same characteristics), the former is a distribution analytically known whereas the latter is a random variable whose distribution can only be known through stochastical experiments -- like Monte-Carlo.

\section{Results\label{sec:B}}

\subsection{Presentation of the different setups}
\par Results of the same problem are given using: first a Monte-Carlo stochastic integration method to solve Equation~\eqref{eq:size-effect:failure-criterion} and second the excursion set theory to solve Equation~\eqref{eq:size-effect:global-stress}.

The stochastic integration provides a full empirical distribution of $\sfailbis(\w)$ where $\sfail(\ps)$ depends directly on the probability parameter $\ps$.
By the definition of the structural failure criterion given by Equation~\eqref{eq:size-effect:global-stress}, both distributions are directly linked, the $n$-quantile of $\sfailbis(\w)$ corresponding to the safety probability $\ps=1-1/n$.
For sake of clarity, both distributions are noted $\sfail$ and are detailed through $\ps$.

\par
\begin{figure}[h!]
  \centering
  \subfigure[Size effect for different safety probability $\ps$ with $\logvar=10$.\label{graph:size-effect:results:p}]{\resizebox{\textwidth}{!}{\begin{tikzpicture}
  \begin{loglogaxis}[
      xlabel={Scale ratio $\rlinv=\Msize/\Lc$},
      ylabel={Normalized failure stress $\sfail$},
      ymin=0.1,
      ymax=1.0,
      xmin=0.001,
      xmax=1000000000,
      ytick={ 0.1, 1 },
      yticklabels={ 0.1, 1 },
      legend entries={$\ps=50\%$,$\ps=90\%$,$\ps=99\%$},
      legend style={at={(0,0)},anchor=south west},
      ]

    \addlegendimage{colorA,line width=2, mark=\markA}
    \addlegendimage{colorB,line width=2, mark=\markB}
    \addlegendimage{colorC,line width=2, mark=\markC}

    \addplot[colorA,line width=2] table[col sep=tab, y expr={0.1*\thisrowno{2}}]{tikz/data/size_effect_1D_m10.0_v10.0_ps0.5.dat};
    \addplot[colorA,line width=2, only marks, mark=\markA] table[y expr={0.1*\thisrowno{5}}]{tikz/data/log_m_10_v_10_q_MonteCarlo.dat};

    \addplot[colorB,line width=2] table[col sep=tab, y expr={0.1*\thisrowno{2}}]{tikz/data/size_effect_1D_m10.0_v10.0_ps0.9.dat};
    \addplot[colorB,line width=2, only marks, mark=\markB] table[y expr={0.1*\thisrowno{4}}]{tikz/data/log_m_10_v_10_q_MonteCarlo.dat};

    \addplot[colorC,line width=2] table[col sep=tab, y expr={0.1*\thisrowno{2}}]{tikz/data/size_effect_1D_m10.0_v10.0_ps0.99.dat};
    \addplot[colorC,line width=2, only marks, mark=\markC] table[y expr={0.1*\thisrowno{2}}]{tikz/data/log_m_10_v_10_q_MonteCarlo.dat};

\end{loglogaxis}

\end{tikzpicture}}}
  \subfigure[Impact of the variance $\logvar$ on the size effect.\label{graph:size-effect:results:v}]                 {\resizebox{\textwidth}{!}{\begin{tikzpicture}
  \begin{loglogaxis}[
      xlabel={Scale ratio $\rlinv=\Msize/\Lc$},
      ylabel={Normalized failure stress $\sfail$},
      ymin=0.1,
      ymax=1.0,
      xmin=0.001,
      xmax=1000000000,
      ytick={ 0.1, 1 },
      yticklabels={ 0.1, 1 },
      legend entries={$\logvar=1$,$\logvar=5$,$\logvar=10$},
      legend style={at={(0,0)},anchor=south west,draw=black},
      ]

    \addlegendimage{colorA,line width=2}
    \addlegendimage{colorB,line width=2}
    \addlegendimage{colorC,line width=2, mark=\markC}

    \addplot[colorA,line width=2] table[col sep=tab, y expr={0.1*\thisrowno{2}}]{tikz/data/size_effect_1D_m10.0_v1.0_ps0.99.dat};

    \addplot[colorB,line width=2] table[col sep=tab, y expr={0.1*\thisrowno{2}}]{tikz/data/size_effect_1D_m10.0_v5.0_ps0.99.dat};

    \addplot[colorC,line width=2] table[col sep=tab, y expr={0.1*\thisrowno{2}}]{tikz/data/size_effect_1D_m10.0_v10.0_ps0.99.dat};
    \addplot[colorC,line width=2, only marks, mark=\markC] table[y expr={0.1*\thisrowno{2}}]{tikz/data/log_m_10_v_10_q_MonteCarlo.dat};

\end{loglogaxis}

\end{tikzpicture}}}
  \caption{Representation of the size effect through a failure stress $\sfail$ estimated over various scales $\rlinv$. The Gaussian related distribution is log-normal of mean $\logesp=10$ and variance $\logvar=1,5$ and $10$ and is based on Gaussian correlated Random Field of correlation length $\Lc=1$. The stochastic integration results (Monte-Carlo) are based on 10\,000 realisations.\label{graph:size-effect:results}}
\end{figure}
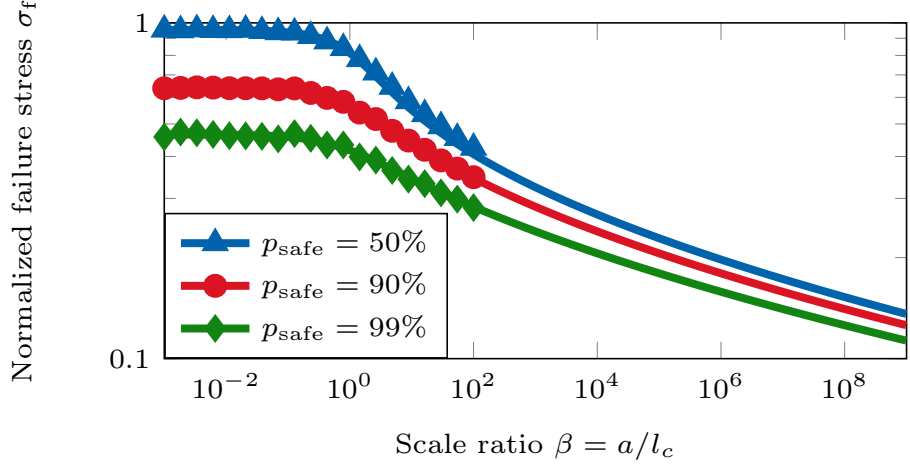
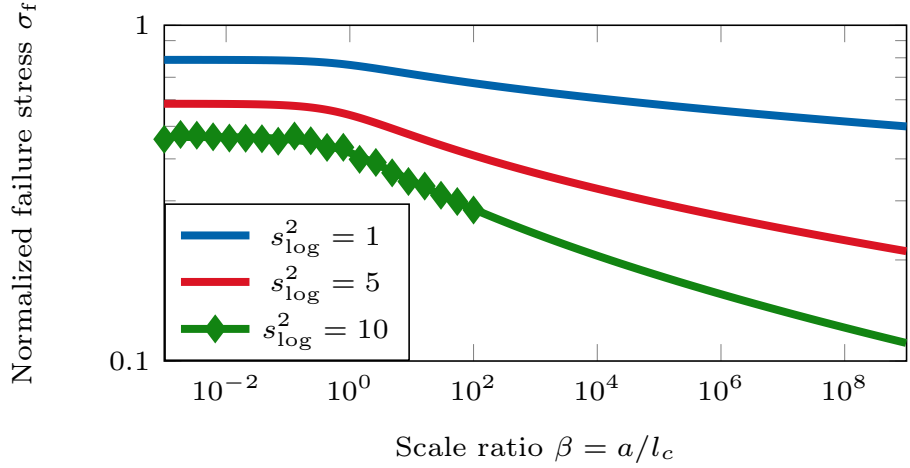

 Figure~\ref{graph:size-effect:results} shows the resulting global failure stresses $\sfail$ as a function of the scale ratio $\rlinv$ with both methods.
The inspection of larger scales is rapidly limited for the Monte-Carlo procedure due to the inconvenient resource consuming aspect of stochastic integrations\footnote{The \textsc{RandomFields} package \citep{schlather_randomfields:_2012} of the \textsc{R} environment \citep{team_r:_2012} has been used in order to do the stochastic integration, using $10\,000$ integration points for each length.}.
For this reason we stopped the computation for scales ratio greater than $10^2$.
On the strength of its analytical base, using the excursion sets theory every scale can be inspected, here for $\rlinv$ varying from $10^{-3}$ to $10^{9}$.

\par Two analysis are made: a first depicted in Figure~\ref{graph:size-effect:results:p}, where the global failure stress is given for various safety probability $\ps=99, 90$ and $50\%$ and a log-normal variance $\logvar=10$. A second in Figure~\ref{graph:size-effect:results:v} for various log-normal variances $\logvar=10, 5$ and $1$ and for a safety probability of $99\%$.
Both are made with a log-normal mean of $\logesp=10$.
In order to link Gaussian and log-normal moments, as required in Equation~\eqref{eq:substitution-log-normal}, the following relationship is used:
\begin{equation}
  \left\{
  \begin{aligned}
    \var&=\frac{1}{2}\ln\left(1+\left(\frac{\logstd}{\logesp}\right)^2\right) = \frac{1}{2}\ln\left(1+\logcvar^2\right) \\
    \esp&=\ln(\logesp)-\frac{1}{2}\var = \ln(\logesp)-\frac{1}{2}\ln\left(1+\logcvar^2\right)
  \end{aligned}
  \label{eq:distribution:gauss-log}
  \right.,
\end{equation}
introducing the coefficient of variation $\logcvar=\logstd/\logesp$.

\par As the mean value $\esp$ and the variance $\var$ of the underlying Gaussian field do not possess a direct physical meaning, the mean value $\logesp$ and the variance $\logvar$ of $\slim(x,\omega)$ do.
The mean value is the structural failure stress for small scales. It can be measured using simple tests on small specimens since it corresponds to the material scale. The variance is related to the heterogeneity of the material by indicating the contrast of strength. Thus it affects the decreasing rate of the size effect for large scales.
An interpretation of the curves is proposed in the following section.

\subsection{Interpretation of the results}

\par As expected, no size effect is observed at small scales ($\rlinv < 10^{-2}$).
For $\ps=50\%$, the value of the failure stress corresponds to the log-normal distribution median (that is, due to the skewness of the distribution, a little less than the mean).
As $\rlinv$ grows, the decrease of $\sfail$, which represents the size effect is observed.
As for the role of $\ps$, results show an expected behaviour.
Indeed, for a safety probability of $\ps=90\%$ (meaning a failure probability of $\pf=10\%$), the failure stress is higher than for a safety of $\ps=99\%$ ($\pf=1\%$).

The three curves drawn in Figure~\ref{graph:size-effect:results:v} represent the impact of the variance on size effect for $\ps=99\%$. As the variance can be seen as a description of the mechanical property discrepancy, results show the natural principle that with increasing variations, the values of $\sfail$ get smaller and the drop over scale ratio higher.

\par The authors are aware that a safety probability of $50\%$ lacks of meaning. We show it to point out a limitation of this model linked with the qualitative definition of ``for low values of $\s$'' in Equation~\eqref{eq:size-effect:expected-EC}, which comes from the excursion sets theory. It means that the larger $\rlinv$ is, the more $\s$ needs to be small for the approximation \eqref{eq:size-effect:link-ECandP} to be accurate (more details in \ref{app:forlow} or \citep{adler_new_2008}).
In the present case, the threshold $\s$ is set by the wanted $\ps$, therefore, for increasing $\beta$ or increasing $\ps$, the accuracy of the model decreases.
This is what the authors believe is observed on the curve $\ps=50\%$ for $\beta>1$ where Monte-Carlo integration and excursion sets theory start to diverge, for lack of precision of the latter.

\par Now, focus is made on an interpretation of the model in terms of Weibull modulus and coefficient of variations.

\subsection{Derivation of a Weibull modulus}

\par The Weibull modulus (noted $\wbull$) can be seen as a way to characterise the importance of the size effect for large scales.
It is defined as a power coefficient linking the structural strength and material geometry.
Considering the scale ratio $\rlinv$, the one dimensional relationship reads \citep{quinn_advanced_1990}:
\begin{equation}
  \frac{ {\sfail}_B }{ {\sfail}_A} = \left(\frac{\rlinv_A}{\rlinv_B}\right)^{-\frac{1}{\wbull}}
\end{equation}
It means that, in the present case, the Weibull modulus is minus the slope of the curves represented Figure \ref{graph:size-effect:results}. For consistency, the modulus is always computed, for very large scales, where the size effect is nearly constant at $\rlinv_A=10^9$ and $\rlinv_B=10^8$.

\par The statistical parameter retained to show the evolution of the Weibull modulus is the coefficient of variation $\logcvar=\logstd/\logesp$. It is relevant since injecting \eqref{eq:distribution:gauss-log} into \eqref{eq:substitution-log-normal} gives:
\begin{equation}
    \sublset = \frac{\ln(\lset) - \ln(\logesp)}{\sqrt{\frac{1}{2}\ln(1+\logcvar^2)}} + \sqrt{\frac{1}{2}\ln(1+\logcvar^2)}
\end{equation}
which makes $\sublset$ depends only on $\logcvar$ and the difference between $\ln(\lset)$ and $\ln(\logesp)$. The later dependency explains why a variation of the mean value of the Random Field shifts the Euler characteristic curves but does not affect the solution of Equation~\eqref{eq:size-effect:global-stress}. Thus the effects on the strength of $\logesp$ and $\logstd$ are linked, and only the coefficient of variation is needed in order to assess the effect of the statistical distribution on the failure strength.

\par
\begin{figure}[h!]
  \centering
  {\resizebox{\textwidth}{!}{\begin{tikzpicture}
  \begin{axis}[
    xlabel={Coefficient of variation $\logcvar$},
    ylabel={Weibull modulus $\wbull$},
%    ymode=log,
%    xmin=0,
%    xmax=20,
    tick label style={/pgf/number format/fixed},
]
    \addplot[colorA,line width=2] table[x expr=\thisrowno{0}, y expr={-\thisrowno{1}}] {tikz/data/size_effect_1D_wmodulus_vs_cvar_v1_ps0.9.dat};
    %\addplot[colorB,line width=2, legend entry={$c=0.05$}] table {tikz/data/size_effect_1D_wmodulus_vs_variance_cvar0.05_ps0.9.dat};
    %\addplot[colorC,line width=2, legend entry={$c=0.01$}] table {tikz/data/size_effect_1D_wmodulus_vs_variance_cvar0.01_ps0.9.dat};
  \end{axis}

\end{tikzpicture}}}
  \caption{Evolution of the Weibull modulus as a function of coefficient of variation.}
  \label{graph:size-effect:wbull:cvar}
\end{figure}
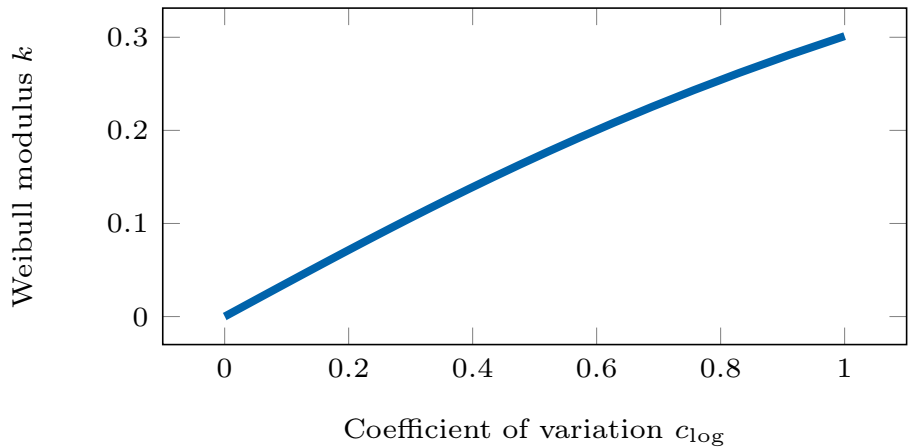

 Finally, Figure \ref{graph:size-effect:wbull:cvar} shows that, as $\logcvar$ decreases, the strength size effect decreases as well.
It is worth noting that this property can be experimentally observed dealing with some classes of materials. This is true for concrete, which exhibits a smaller Weibull modulus when considering higher performance formulations (see \citep{rossi_scale_1994} for experimental results).

%In order to understand why the behavior is

%\begin{figure}[h!]
%  \centering
%  \subfigure[Size effect curves for $\logcvar=1$\label{graph:size-effect:se-vs-cvar}]{\resizebox{0.495\textwidth}{!}{\input{tikz/log_1D_var_std_constant_cvar}}}
%  \subfigure[Variation of $\logvar$ and $\logesp$ for constant $\logcvar=0.1, 0.05$ and $0.01$]{\resizebox{0.495\textwidth}{!}{\input{tikz/log_1D_wmodulus_vs_variance}}}
%  \caption{Impact of the size effect as function of the mean or the variance of the Random Field with constant coefficient of variation.}
%\end{figure}
%
%\begin{figure}[h!]
%  \centering
%  \subfigure[Variation of $\logvar$ for constant $\logcvar$.\label{graph:size-effect:wbull:var}]{\resizebox{\textwidth}{!}{\input{tikz/log_1D_wmodulus_vs_variance}}}
%  \subfigure[Variation of $\logcvar$.\label{graph:size-effect:wbull:cvar}]                      {\resizebox{\textwidth}{!}{\input{tikz/log_1D_wmodulus_vs_cvar}}}
%  \caption{Evolution of the Weibull modulus $\wbull$ as a function of the Random Field statistical properties.\label{graph:size-effect:wbull}}
%\end{figure}

%\begin{figure}[h!]
%  \centering
%  \subfigure[Variation of $\logesp$]{\resizebox{0.495\textwidth}{!}{\input{./tikz/log_1D_wmodulus_vs_mean}}}
%  \subfigure[Variation of $\logvar$]{\resizebox{0.495\textwidth}{!}{\input{./tikz/log_1D_wmodulus_vs_variance}}}
%  \caption{Impact of the size effect as function of the mean and the variance of the Random Field.}
%\end{figure}

\section{Conclusion\label{sec:con}}

A theoretical scaling law for the size effect of the strength of brittle materials has been proposed.
The key idea is to try to link the intrinsic heterogeneous geometry of those materials to the macroscopic strength of a structure of a given size $a$.
In order to represent this heterogeneous character, we have used correlated Random Fields that, thanks to their spatial structure, may be used to set a ``material scale'' by opposition to the ``structural scale'' $a$.
Moreover, using quite recent results from \citep{adler_random_2007} on the geometry of excursion sets, one can analytically compute the probability of exceedance of Random Fields and thus compute the whole probability distribution of the structural strength.
Having this distribution in hands and chosing for a given risk (ie. probability failure), it is straightforward to calculate the structural strength.

Although covering a large range of sizes and showing excellent agreement with experimental considerations, this scaling law is unfortunately restricted to 1D tension.
Still some extensions to both 3D structures and loading paths are possible.
Those extensions may be based on theoritical results for the geometry of 3D excursions sets.
More precisely, one can add more information dealing with the geometry of the material and of the failure process zone, \ie its geometry (volume and surface) as well as its topology (percolation probability).

Finally, we shall progress to the definition of an identification procedure for some specific materials. Obviously such a procedure would need to be based on experimental results, for example a family of similar tests (simple compression, bending, ...) on homothetical structures of growing sizes.

\appendix
\section{For ``low'' values of the thresholds\label{app:forlow}}
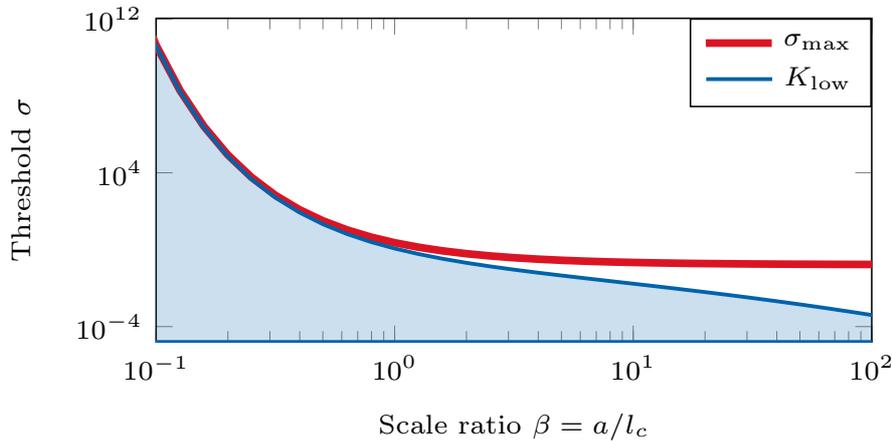
\begin{figure}[h!]
  \centering
  \resizebox{\textwidth}{!}{\begin{tikzpicture}[]

  \begin{loglogaxis}[
      xlabel={Scale ratio $\rlinv=\Msize/\Lc$},
      ylabel={Threshold $\lset$},
      x tick label style={/pgf/number format/fixed},
      legend style={at={(1,1)},anchor=north east},
      xmin=0.1,
      xmax=100,
      log origin=infty,
    ]
    \addplot[colorB,line width=2, legend entry={$\lsetmax$}] table[y expr={\thisrowno{1}}]{tikz/data/size_effect_1D_max_m0.5_v2_function.dat};
    \addplot[colorA, fill, fill opacity=0.2, line width=1, legend entry={$\lsetlow$}] table[y expr={\thisrowno{3}}]{tikz/data/size_effect_1D_max_m0.5_v2_function.dat} \closedcycle;
  \end{loglogaxis}

\end{tikzpicture}}
  \caption{Threshold maximizing the Euler characteristic $\lset_\text{max}$ and the corresponding set of ``low'' thresholds $\lsetlow$. The Gaussian related distribution is log-normal and is based on Gaussian correlated Random Field of mean $\esp=0.5$, variance $\var=2$ and correlation length $\Lc=1$. The hitting set is cumulative $\HS=\zlset$.\label{graph:ECs:maximums}}
\end{figure}

An analysis of Equation~\eqref{eq:size-effect:expected-EC} giving the expected Euler characteristic shows that its maximum is always defined at $\lset_\text{max}=\exp(\sqrt{\pi}\std/\rlinv+\esp)$ (see Figure~\ref{graph:ECs:maximums})). However, an interesting result is the shifting of the threshold corresponding to this maximum since it qualitatively describes the term ``for low values of $\lset$'' corresponding to the appearance of the first connected component, ``low'' being lower than $\lset_\max$. In other words, as $\rlinv$ increases as ``low'' corresponds to lower values and, reciprocally, as $\rlinv$ decreases, as ``low'' can means large values. It can be defined by the set:
\begin{equation}
  \lsetlow=\left\{\lset\ |\ \lset\le\lfactor\lsetmax\ \text{with}\ 0\leq\lfactor\leq1, \frac{\partial\lfactor(\rlinv)}{\partial\rlinv}<0, \lim_{\rlinv\to0}\lfactor(\rlinv)=1,  \lim_{\rlinv\to\infty}\lfactor(\rlinv)=0\right\}.
  \label{eq:relation-low-max}
\end{equation}
Only for sake of graphic depiction, Figure~\ref{graph:ECs:maximums} shows $\lsetlow$ for $\lfactor$ arbitrary taken to be:
\begin{equation}
\lfactor(\rlinv)=\frac{1}{2}\left(1-\frac{2}{\sqrt{\pi}}\int_0^{\log(\rlinv)} e^{-t^2}dt\right),
\end{equation}
thus fulfilling the necessary conditions of Equation~\eqref{eq:relation-low-max}. It is reminded that this threshold is qualitatively drawn. Actual limitations are discussed in the last section of this paper when the theory is confronted with numerical simulations.

\bibliographystyle{plainnat}
\bibliography{includes/biblio}

\end{document}